\documentclass[12pt]{article}
\usepackage{fleqn}
\def\be{\begin{equation}}
\def\ee{\end{equation}}
\def\ba{\begin{eqnarray}}
\def\ea{\end{eqnarray}}

\def\fun#1#2{\lower3. 6pt\vbox{\baselineskip0pt\lineskip.
9pt\ialign{$\mathsurround=0pt#1\hfill##\hfil$\crcr#2\crcr\sim\crcr}}}

\usepackage{graphicx}

\begin{document}
\begin{titlepage}
\null\vspace{-62pt}

\vspace{0.5in}
\centerline{\large \bf{Gravitational Collapse of Dark Energy Field Configurations}}

\vspace{.5in}
\centerline{Anupam Singh}
\vspace{.4in}
\centerline{{\it Physics Department,  L.N. Mittal I.I.T,
Jaipur, Rajasthan, India.}}
\vspace{.5in}
\baselineskip=24pt
\begin{quotation}

It is now a well established fact based on several observations that we live in a Universe dominated by dark energy. 
The most natural candidates for dark energy are fields in curved space-time. 
We develop the formalism to study the gravitational collapse of fields given any general potential.
We apply this formalism to models of dark energy motivated by particle physics considerations.
We solve the resulting evolution equations which determine the time evolution of field configurations as well as
the dynamics of space-time.

\end{quotation}
\end{titlepage}
\newpage

\baselineskip=24pt

\vspace{24pt}

\section{Introduction}

Increasing accuracy in astronomical observations has  led us to an
increased precision in the determination of cosmological parameters.
This in turn led us to critically re-examine our cosmological models.
In particular,  the accurate determination of the Hubble constant and
the independent determination of the age of the universe forced us to
critically re-examine the simplest cosmological model
- a flat universe with a zero cosmological constant\cite{Pierce,Freedman}.
These observations forced the current author\cite{Singh} to consider the idea of
a small non-vanishing vacuum energy due to fields as playing an important role in the Universe.
Subsequently there has been a large body of work both on the observational and theoretical side
that has firmed up our belief in what we now call dark energy.

Given that we now believe that dark energy is the dominant component of the Universe, it is now a pressing need
to understand the dynamics of dark energy and in particular the gravitational dynamics of dark energy.
In this Letter we describe the gravitational dynamics of dark energy and show the 
gravitational collapse of dark energy field configurations.

Before we start our detailed look at the gravitational dynamics of dark energy field configurations we would like to
quickly introduce field theory models of Dark Energy so that all readers can readily relate to the discussion that follows.

Discussions of field theory models for dark energy and connections to particle physics were discussed in detail earlier by the current author\cite{Singh}.
It was noted in that article that these fields must have very light mass scales in order to be cosmologically relevant today.
Discussions of realistic particle physics models with particles of light masses capable of generating interesting cosmological consequences
have been carried out by several authors\cite{GHHK}. It has been
pointed out that the most natural class of models in which to realise these ideas are models of neutrino masses with 
Pseudo Nambu Goldstone Bosons (PNGB's). The reason for this is that the mass scales associated with such
models can be related to the neutrino masses, while any tuning that needs to be
done is protected from radiative corrections by the symmetry that gave rise
to the Nambu-Goldstone modes\cite{'thooft}.

Holman and Singh\cite{HolSing} studied the finite temperature behaviour of the
see-saw model of neutrino masses and found phase transitions in this model
which result in the formation of topological defects.  In fact, the critical
temperature in this model is naturally linked to the neutrino masses.

In our next section describing the formalism and evolution equations for studying the
gravitational dynamics of dark energy field configurations we will first write down the equations for
fields with any general potential to keep the initial discussion general. However, in light of the above discussion
on realistic models with light masses we will soon specialize to PNGB models 
which have a potential with simple and explicit form. Thus, for this purpose we will take the simplest PNGB potential\cite{GHHK}.

\section{The Evolution Equations}

We wish to study the gravitational dynamics of Dark Energy field configurations.
The dynamics of fields in cosmological space-times has been extensively discussed
elsewhere (see e.g. Kolb and Turner\cite{rockymike} ). Likewise, gravitational collapse
in the context of general relativity has also been extensively discussed elsewhere 
(see e.g. Weinberg\cite{weinberggr}). These ideas can be pulled together to write 
down the evolultion equations describing the coupled dynamics of the field configurations and
space-time interacting with each other. This is what we will do in this section. In the next section we
will discuss the solutions to these evolution equations.

To study gravitational collapse in a cosmological setting we can start with a metric of the form:

\be
ds^2 = dt^2 - U(r,t) dr^2 - V(r,t) \left[   d\theta^{2} + 
\sin^{2}\theta d\phi^{2} \right]
\ee

where,  $ (t, r, \theta, \phi) $
are the comoving coordinates describing a space time point. 
This metric, is of course a generalization of the usual FRW metric used to study cosmological space-times.
Note that the functions $U(r,t)$ and $V(r,t)$ are functions of both space and time and can capture both homogeneous
cosmological expansion as well as inhomogeneous gravitational collapse under appropriate circumstances.

We are interested in studying the dynamics of fields in space-time. For this purpose and to keep the initial discussion
as general as possible we will consider a field $\Phi$ with a Lagrangian $L$ given by

\be
L = \frac{1}{2} \partial^\mu \Phi \partial_\mu \Phi - {\cal V}(\Phi)
\ee

where ${\cal V}$ is the potential for the field $\Phi$ and is for now a general function.
Later, when we consider the physically motivated PNGB models this potential will take on 
a specific functional form.

By using standard techniques one can then write
down the evolution equations for the metric and the field. These can be written down in the form:

\be
\ddot V =  2 \left[ -1 + \frac{ V^{\prime \prime}}{2 U}  -  \frac{ V^{\prime} U^{\prime}}{4 U^2} - \frac{ \dot V \dot U}{4 U} + 8 \pi G V \left( \frac{\rho}{ 2} -\frac{ P}{ 2}- \frac{(\Phi^{\prime})^2 }{3 U} \right) \right ]
\ee

\be
\ddot U =  2  U \left[ - \frac{ \ddot V}{V}  + \frac{ \dot U ^2}{4 U^2} + \frac{ \dot V ^ 2}{2 V^2} - 4 \pi G  \left( \rho + 3  P \right) \right ]
\ee

\be
\ddot \Phi =   \frac{ \Phi^{\prime \prime}}{U}  -  \dot \Phi \left[\frac{\dot V}{V}+\frac{ \dot U}{2 U} \right] + \frac{\Phi^{\prime} }{U} \left[ \frac{V^\prime}{V} - \frac{U^\prime}{2 U}\right] - \frac{\partial {\cal V} (\Phi)}{\partial \Phi}
\ee

where a dot represents a partial derivative w.r.t. $t$ and a prime represents a partial derivative w.r.t. $r$. Further, 

\be
\rho = \frac{1}{2} \dot \Phi ^2 + {\cal V} (\Phi) +  \frac{(\Phi^{\prime})^2 }{2 U}
\ee

and

\be
P =  \frac{1}{2} \dot \Phi ^2 -  {\cal V} (\Phi) +  \frac{(\Phi^{\prime})^2 }{6 U} .
\ee

The above equations are true for any general potential ${\cal V} (\Phi)$. 
One can of course write down the corresponding equations for PNGB fields.
The simplest potential one can write down for the physically motivated PNGB fields \cite{GHHK} can be written in the form:

\be
{\cal V}(\Phi) = m^4 \left[ K - \cos(\frac{\Phi}{f} ) \right]
\ee

As discussed in \cite{GHHK} m is of order the neutrino mass and K is of order 1. We will consider such a potential for studying the dynamics in the next section.

\section{Solution to the Evolution Equations}

The evolution equations described above can be solved numerically to study the gravitational collapse of the field configurations. The key issue we want to understand and address is the timescale on which the gravitational collapse happens. Clearly if this timescale for collapse is equal to or larger than the age of the Universe then the gravitational collapse has no practical significance or implications. On the other hand, if the gravitational collapse happens on a timescale shorter than the age of the Universe then we must consider the gravitational collapse of the dark energy field configurations.

Guided by the evolution equations given in the previous section we define dimensionless quantitites such that the field is measured in units of $f$ and time and space are measured in units of $\frac{f}{m^2}$.

\begin{figure}
\centering
\includegraphics{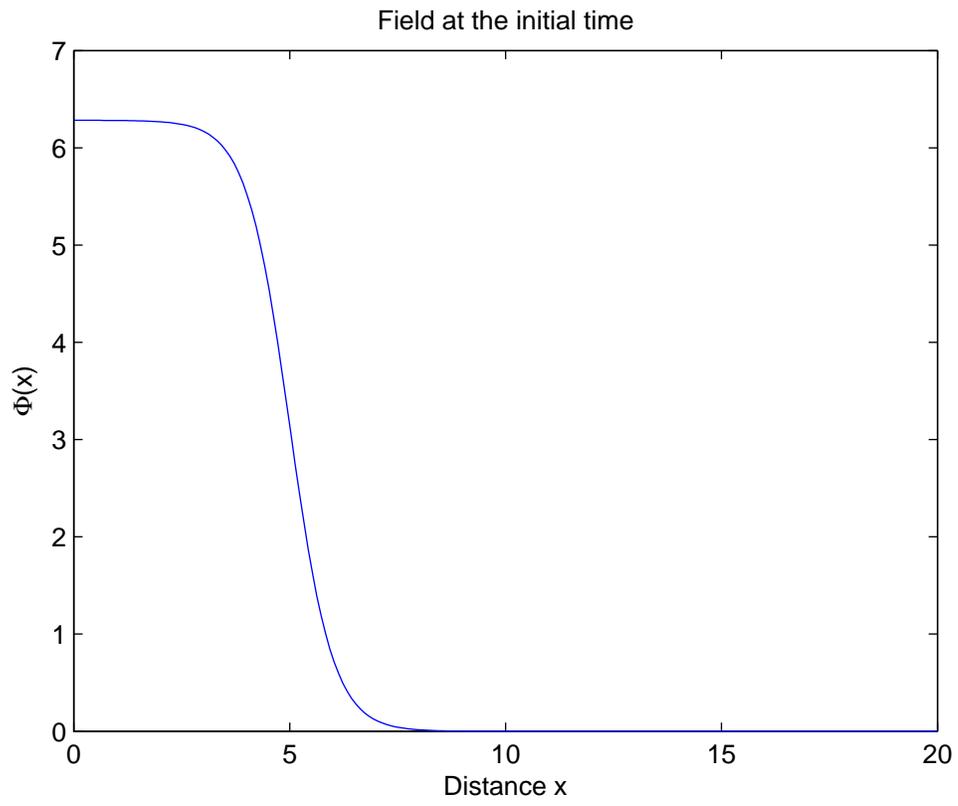}
\caption{Initial Field configuration} \label{Initial Field}
\end{figure}

The initial field configuration in terms of these dimensionless valiables is given in the figure \ref{Initial Field} that follows.

\begin{figure}
\centering
\includegraphics{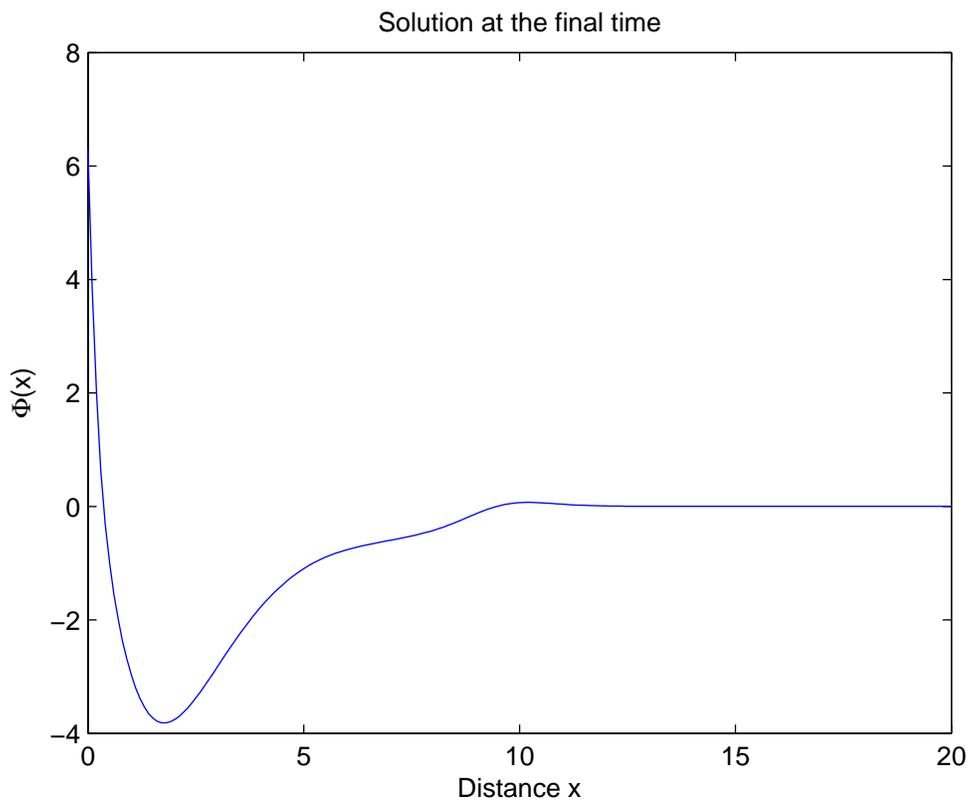}
\caption{Final Field configuration} \label{FinalField}
\end{figure}

The final field configuration in terms of these dimensionless valiables is given in figure \ref{FinalField}.

\begin{figure}
\centering
\includegraphics{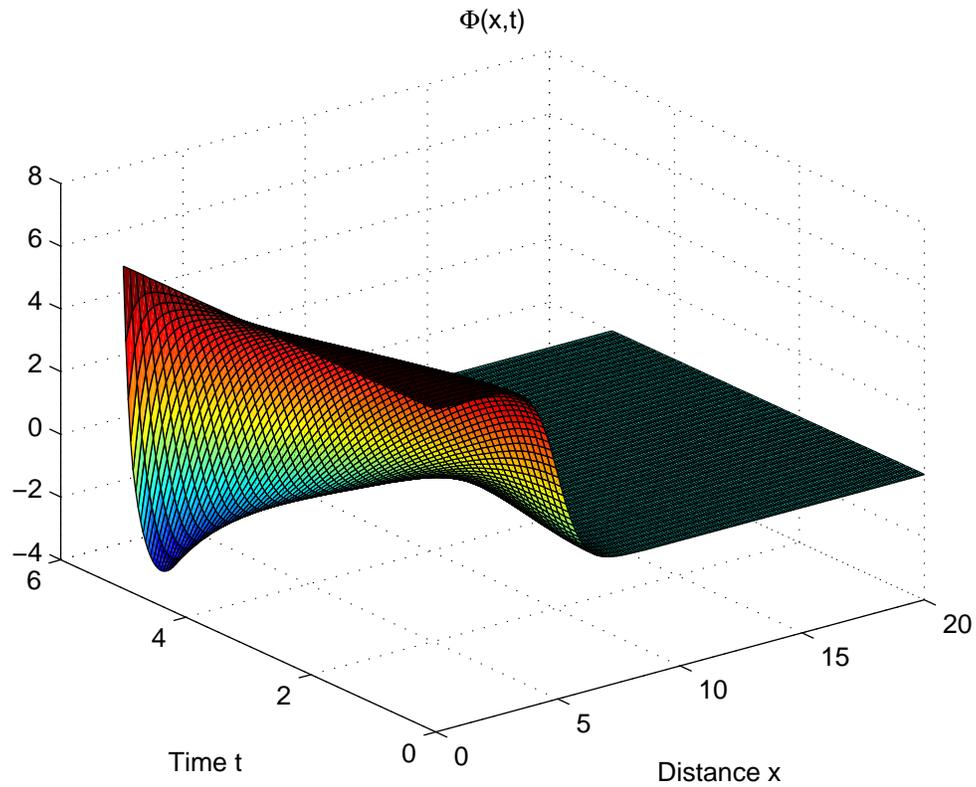}
\caption{Field configuration in space-time} \label{SpaceTimeField}
\end{figure}

From this it can be clearly seen that field configuration has collapsed and the timescale for collapse can be seen by studying the figure \ref{SpaceTimeField}. Since the units of time are given by $\frac{f}{m^2}$, we note that gravitational collapse happens on timescales of $ \sim \frac{f}{m^2}$. This timescale is much shorter than the age of the Universe.

\newpage

\section{Summary and Conclusions}

It is now a well established fact that the Universe is dominated by Dark Energy. The most promising candidate for Dark Energy is the energy density due to fields. Understanding the dynamics of these Dark Energy fields is of paramount importance. In this Letter we developed the formalism to study the gravitational collapse of fields given any general potential for the fields. We then applied this formalism to the simplest and most natural model of dark energy fields motivated by Particle Physics considerations. We reported on the results obtained by numerically solving the resulting evolution equations. Our results show that gravitational collapse of dark energy fields happens on a timescale which is short compared to the age of the Universe.


\centerline{\bf ACKNOWLEDGEMENTS}

This work was supported in part by a research grant from L.N.Mittal I.I.T.

\frenchspacing


\newpage

\end{document}